\documentclass[%
 reprint,
 amsmath,amssymb,
 aps,
]{revtex4-1}

\usepackage{graphicx}
\usepackage{dcolumn}
\usepackage{bm}
\usepackage{natbib}
\begin{document}


\title{Visualization of Open and Closed String Worldsheets. Part I: Software Development}

\author{Graham W. Van Goffrier}
 \email{graham.van@maine.edu}
\author{Neil F. Comins}%
 \email{galaxy@maine.edu}
\affiliation{%
 Department of Physics and Astronomy, University of Maine
}%


\date{\today}

\begin{abstract}
In bosonic string theory, the solutions to the string equations of motion may be expressed as two-dimensional manifolds in a relativistic spacetime. We develop MATLAB software for the generation of open and closed string solutions and for the convenient visualization of these solutions. 

\end{abstract}

\pacs{Valid PACS appear here}
\maketitle


\section{\label{sec:level1}Introduction}


String theory, at its most familiar conceptual level, is the study of one-dimensional “strings” that vibrate and move through spacetime according to relativistic principles. This motion can be said to trace out a surface in spacetime, called a worldsheet, which contains all positions of a string throughout time. The equations that govern the motion of strings may be derived, as shown in \cite{zwiebach}, from a reasonable assumption that the local relativistic area measured on the worldsheet be minimized. The term ‘area’ is used liberally here to describe the two-dimensional manifold in spacetime that is swept out by the string position throughout time.

In this paper we present the equations that describe relativistic string behavior, and the development and testing of a program which enables us to visualize this behavior. In order to focus on representing non-quantum spacetime dynamics, this work did not seek to explore the more recent and more realistic string theories, such as superstring theory and M-theory which dominate discourse in the field. Classical relativistic strings were considered exclusively, analogous to a relativistic point particle such as an electron, but without inclusion of probabilistic quantum effects.  It should further be noted that the light-cone gauge string solution technique, detailed in Section II.A, is a significant step towards one string quantization paradigm (known as light-cone gauge quantization) that reflects important results from modern string theory such as graviton states.

The mode expansion technique of string solution introduced in Chapter 9 of \cite{zwiebach}, allows for the development of MATLAB software for the generation of string solutions.  This software calculates the Virasaro mode coefficients given input parameters and mode coefficients for transverse oscillations, and generates data points for the position functions of the worldsheet, which are plotted. After an overview of relevant theory in Section II, these developments are detailed in Section III alongside examples for demonstration of functionality. More physical aspects of examples are detailed in a companion paper \cite{partii}.

\section{Theory}

The Nambu-Goto action action $\mathbf{S_{NG}}$, a functional of each possible trajectory in some environment, is a path integral of the Lagrangian for the system:

\begin{equation}
\label{eq:ngaction}
\mathbf{S_{NG}} = \frac{1}{2\pi\alpha'}\int d\tau\,d\sigma\,\sqrt{(\dot{X}\cdot X')^{2} - (\dot{X})^{2}(X')^{2}}.
\end{equation}

$\tau$ and $\sigma$ are linearly independent parameters of the string worldsheet, representing evolution of the string and the length along the string respectively. X is a vector function giving the spacetime location of the worldsheet as a function of $\tau$ and $\sigma$, with derivatives denoted by $\dot{X}$ and $X'$ respectively. $\alpha'$ is the slope parameter, a commonly used expression of the single free constant of string theory which must be specified rather than derived from other principles. In this paper the speed of light c is set equal to 1 (unitless) for the sake of descriptive ease. The above action can be shown to be proportional to the relativistic area of the string's worldsheet \cite{zwiebach}, setting string theory in the familiar context of a variational problem.

A very important property \cite{zwiebach} of the Nambu-Goto action is that it is reparameterization invariant; the parameters $\tau$ and $\sigma$ may be selected as desired so long as they remain linearly independent on the worldsheet manifold. The chosen parameterization for $\tau$ is called the static gauge, and simply sets $\tau$ equal to time. The $\sigma$-parameterization is then selected so that 

\begin{equation}
\label{eq:cons1}
\frac{dX}{d\tau}\cdot\frac{dX}{d\sigma}=0.
\end{equation}

This parameterization is also chosen so that the energy density of the string is constant along sigma, and for this reason is sometimes called the energy parameterization. In this setup, the equations of motion derived from the action \cite{zwiebach} reduce to the wave equation:

\begin{equation}
\label{eq:wave1}
\frac{d^2X}{d\sigma^2}-\frac{1}{c^2}\frac{d^2X}{d\tau^2}=0
\end{equation}

with additional parameterization constraint:

\begin{equation}
\label{eq:cons2}
\bigg(\frac{dX}{d\sigma}\bigg)^2+\frac{1}{c^2}\bigg(\frac{dX}{d\tau}\bigg)^2=1.
\end{equation}

Together with \eqref{eq:cons1}, the constraint \eqref{eq:cons2} may be rewritten as:

\begin{equation}
\label{eq:constogether}
\bigg(\frac{dX}{d\sigma}\pm\frac{1}{c^2}\frac{dX}{d\tau}\bigg)^2=1.
\end{equation}

General solutions of the above wave equation for open strings with free boundary conditions take the form:

\begin{equation}
\label{eq:opengen}
X(\tau,\sigma)=\frac{1}{2}[F(ct+\sigma)+F(ct-\sigma)]
\end{equation}

where $F$ is known to be quasiperiodic (periodic with constant translation), and $\dot{F}$ is known to be a unit vector. Similarly, for closed strings, general solutions take the form 

\begin{equation}
\label{eq:closedgen}
X(\tau,\sigma)=\frac{1}{2}[F(ct+\sigma)+G(ct-\sigma)]
\end{equation}

where both $F$ and $G$ have the same properties of quasiperiodicity and unit vector derivative. $F$ and $G$ may be considered as independent left- and right-moving waves on the closed string; an open string results when these waveforms coincide.

\subsection{Light-Cone Gauge Mode Expansion}

Mode expansion refers to the general technique of solving differential equations by expressing the solution as a sum of complex-exponential functions with coefficients to be selected, similar to a Fourier series expansion. This technique was found to be very useful for the generation of a general class of string solutions.

Beginning with the general open-string solution form of \eqref{eq:opengen}, the mode expansion is derived \cite{zwiebach} to be:

\begin{equation}
\label{eq:xmudef}
X^{\mu}(\tau,\sigma) = x^{\mu}_{0} + \sqrt{2\alpha'}\alpha^{\mu}_{0}\tau + i\sqrt{2\alpha'}\displaystyle\sum_{n\neq0} \frac{1}{n}\alpha^{\mu}_{n}e^{-in\tau}cos(n\sigma)
\end{equation}

$x^{\mu}_{0}$ represents the initial string position, and $\alpha^{\mu}_{0}$ represents the initial string momentum, scaled to be a unitless quantity. All the $\alpha^{\mu}_{n}$ are complex coefficients of the string oscillatory modes. These are also defined in such a way that they are unitless, for which reason the $\sqrt{2\alpha'}$ coefficient, which has units of inverse energy \cite{zwiebach}, is included. To fully determine the general string solution in terms of the $\alpha^{\mu}_{n}$ mode coefficients, the constraint given in \eqref{eq:cons2} must also be expressed in terms of the mode expansion. This is shown in Equation \eqref{eq:cons2} to be:

\begin{equation}
\label{eq:consnew}
\bigg(\frac{d\vec{X}}{d\sigma}\pm\frac{1}{c^2}\frac{d\vec{X}}{d\tau}\bigg)^2=2\alpha'\bigg(\displaystyle\sum_{n \in \mathbb{Z}} \alpha^{\mu}_{n} e^{-i n(\tau\pm\sigma)}\bigg)^2=1.
\end{equation}

The number of cross-terms of this equation goes as $n^2$, and proves very unwieldy for efficiently producing string solutions. Turning to the Light-Cone Gauge (LCG) formalism for string solutions eliminates this constraint by incorporating it into the mode coefficients which make up the solution terms.

The light-cone coordinates, simply a rotated coordinate frame for Minkowski space, are defined by:

\begin{equation}
\label{eq:lcrelate}
x^{+} = \frac{1}{\sqrt{2}}(x^{0}+x^{1})
\qquad
and
\qquad
x^{-} = \frac{1}{\sqrt{2}}(x^{0}-x^{1}).
\end{equation}

To construct the LCG for string solutions, new parameterizations for both $\tau$ and $\sigma$ are defined as follows:

\begin{equation}
\label{eq:lcdef}
X^{+}(\tau,\sigma) = \beta\alpha'p^{+}\tau
\qquad
and
\qquad
p^{+}\sigma = \frac{2\pi}{\beta}\int_0^{\sigma}d\tilde{\sigma}\mathcal{P}^{\tau+}(\tau,\tilde{\sigma})
\end{equation}

where $\beta$ is a constant that is set to 1 for closed strings or 2 for open strings, and ${P}^{\tau+}$ may be interpreted as energy-momentum density in the $x^{+}$ direction. As in the static gauge \eqref{eq:cons1}, $\tau$ is inherently linear, while $\sigma$ is proportional to a cumulative energy density along "slices" of the worldsheet. The distinction is that these slices are now orthogonal to the $x^{+}$ direction rather than the $x^0$ direction.

Applying \eqref{eq:lcrelate} and \eqref{eq:lcdef} to the constraint given in \eqref{eq:cons2}, the following simplification is derived \cite{zwiebach}:

\begin{equation}
\label{eq:consLC}
\bigg(\frac{dX^{-}}{d\sigma}\pm\frac{1}{c^2}\frac{dX^{-}}{d\tau}\bigg)=\frac{1}{\beta\alpha'}\frac{1}{2p^{+}}\bigg(\frac{dX^{I}}{d\sigma}\pm\frac{1}{c^2}\frac{dX^{I}}{d\tau}\bigg)^2
\end{equation}

All $X^+$ terms have vanished from this constraint due to the chosen proportionality of $X^+$ to $\tau$ by \eqref{eq:lcdef}, fully specified by $p^{+}$. Therefore, by fixing $p^{+}$ and $X^{I}$, all $\frac{dX^{-}}{d\sigma}$ and $\frac{dX^{-}}{d\tau}$ are determined, so determining $X^{-}$ up to initial position $x^{-}_0$. Thus any selection of the transverse mode coefficients $\alpha^{\mu}_{n}$ must produce a valid string solution. Also, as the transverse coordinates are not changed in LCG as compared to the static gauge, their mode-expansion remains identical \eqref{eq:xmudef}:

\begin{equation}
\label{eq:xIdef1}
x^{I}(\tau,\sigma) = x^{I}_{0} + \sqrt{2\alpha'}\alpha^{I}_{0}\tau + i\sqrt{2\alpha'}\displaystyle\sum_{n\neq0} \frac{1}{n}\alpha^{I}_{n}e^{-in\tau}cos(n\sigma)
\end{equation}

However, the mode coefficients of $X^{-}$ are no longer independent, due to the constraint stated above \eqref{eq:consLC}. These mode coefficients, $L_{n}^{\perp}$, instead must be expressed in terms of the transverse mode coefficients and $p^{+}$, and in this form are known as the Virasaro modes:

\begin{equation}
\label{eq:virmode}
L_{n}^{\perp}=\frac{1}{2}\displaystyle\sum_{p} \alpha_{p}^{I}\alpha_{n-p}^{I}
\end{equation}

where the indices $I$ are summed over. Then the functional form of $X^{-}$ may be written as:

\begin{equation}
\label{eq:xndef1}
X^{-}(\tau,\sigma) = x^{-}_{0} + \frac{1}{p^{+}}L^{\perp}_{0}\tau + \frac{i}{p^{+}}\displaystyle\sum_{n\neq0} \frac{1}{n}L^{+}_{n}e^{-in\tau}cos(n\sigma)
\end{equation}

The corresponding mode expansion for closed strings is also derived in \cite{zwiebach}, with the only significant change in structure being the presence of two full sets of modes, denoting left- and right- moving solutions.

\section{\label{sec:level1}MATLAB Implementation}

As has been described, all possible string worldsheets may be described by a single set of coordinate expressions, once a finite set of parameters and countably infinite set of mode expansion coefficients are specified. Because of the lack of constraints upon these parameters (any real value may be specified for the parameters, and any complex value for the mode expansion coefficients), this method of generating solutions lends itself well to software implementation.

MATLAB was selected as the development platform due to its high potential for parallelizability. In addition, MATLAB has high-level plotting and visual effects capacity, which are valuable in the visualization of generated string worldsheets.

The MATLAB script produced is referred to as the Light-Cone Gauge Euclidean Minimal Surface Generator (LEGen), as the original motivation of this work was the investigation of minimal surfaces in Euclidean space by applying string methods. The design of the underlying algorithms, as well as code debugging, will now be discussed in order of the flow of the LEGen script.

\subsection{LEGen Design and Debugging}

LEGen is modular in design; distinct scripts provide for the generation of open and closed strings from chosen mode-coefficients. Additionally, string solutions may be generated by inputting a general functional form of the left and right travelling waves, a feature which will be relevant to the study of string cusps and kinks in Paper II: Examples and Applications. All these scripts make use of a shared body of functions, both for calculating worldsheet data from inputs and for visualizing the resultant worldsheets.

\subsubsection{Parameter Extraction}

In order to facilitate user-input, an efficient and readable formatting for input text files was constructed. The required parameters are the dimensionality $D$, light-cone energy $p^+$, initial positions $x^{-}_{0}$ and $x^{I}_{0}$, and transverse mode coefficients $\alpha^I_n$. These are stored row-by-row in the text file, represented as ordered pairs of polar coordinates for each complex value. For $D>3$ string calculations, the higher-dimensional vector for $\alpha^I_n$ is represented by space-separated ordered pairs in the same row. An example solution for a rotating straight string in three space dimensions is described by the following input file:

\begin{verbatim}
4               dimensionality        
(1,0)           p+ momentum
(0,0)           x- initial position
(0,0)           xI transverse initial position
(0,0) (0,0)     aIn transverse modes    n=0  
(1,0) (1,0)                             n=1
(1,0) (1,0)                             n=2
\end{verbatim}

This format conveniently allows for in-line comments without corrupting the input, which aid in easy interpretation of the input parameters by users.

\subsubsection{Virasaro Mode Calculation}

As detailed in Section II.A, the transverse mode coefficients are sufficient, in the light-cone gauge, to describe all oscillatory string motion. The Virasaro modes $L_{n}^{\perp}$ of the $x^{-}$ direction may be calculated explicitly from the transverse modes $\alpha^I_{n}$ \eqref{eq:virmode}.

An arbitrary string solution may have an infinite number of nonzero transverse modes. However for the purposes of this paper, we assume that only a finite set of modes are nonzero. As in Fourier analysis this assumption serves as a low-pass filter, excluding solutions with rapid oscillatory terms. A variety of simplifications are set forth in \cite{krijn_mode} and \cite{kraai_mode}, which identify cases wherein the properties of the transverse modes eliminate some terms of the above definition. These ideas were expanded into a functioning algorithm, which forms the centerpoint of this MATLAB script.

One essential step is the calculation of a limit to values of p which must be considered in \eqref{eq:virmode} to ensure all nonzero $\alpha^I_{n}$ are accounted for. Assuming that there is some value m for which:

\begin{equation}
\label{eq:mlimit}
\forall n \:|\: (\mid n \mid \:> m \: \& \: n \in \mathbb{Z}) : \:\: \alpha_{n}^{I} = 0.
\end{equation}

From this, it was derived that to calculate the nth Virasaro mode $L_{n}^{\perp}$, p in the range:

\begin{equation}
\label{eq:prange}
\frac{n}{2} - limit \leq p \leq \frac{n}{2} + limit
\end{equation}
need only be considered, where $limit$ is defined by:

\begin{equation}
\label{eq:limit}
limit = m - \frac{\mid n \mid}{2} .
\end{equation}

\subsubsection{Numerical Worldsheet Calculation and Output Plots}

With the Virasaro modes identified, spacetime values for the string worldsheet in all light-cone coordinates may be calculated by \eqref{eq:lcdef}, \eqref{eq:xndef1}, and \eqref{eq:xIdef1}. Applying the light-cone coordinate relations \eqref{eq:lcrelate} to the resultant arrays, the worldsheet is expressed in standard Minkowski coordinates.

For a quantity of $\tau$ and $\sigma$ values set by the user, where $\tau$ may cover any desired real domain, and $\sigma$ customarily ranges between 0 and $2\pi$, LEGen calculates the coordinates of the string worldsheet using the above equations.

Plotting is then performed for a given worldsheet by MATLAB's native "surf" function, which generates a three-dimensional colorized and shaded surface. This surface may be rotated and otherwise manipulated by the viewer for added ease of use, and is set to include grid lines on the surface between plotted points. These grid lines are a very valuable feature, as they correspond to the equivalence curves of $\tau$ and $\sigma$.

Key visualization tools which were implemented include animation of string propagation over time, plotting of arbitrary worldsheet sub-regions for more careful analysis, and tagging of string regions of physical interest. In particular, non-differentiable points on the string known as cusps and kinks are identified automatically, and are explored in Paper II: Examples and Applications.

\subsection{Demonstration Examples}

A diverse selection of test input parameters were run on LEGen, both as a part of the debugging process and for demonstrating various aspects of functionality. Open-string examples which were key to confirming proper operation of the model are given in this subsection; more thorough and physically-interesting examples are developed in Paper II of this work, and demonstrate the full capabilities of this software.

\begin{figure}[h!]
    \centering
    \includegraphics[width=0.5\textwidth]{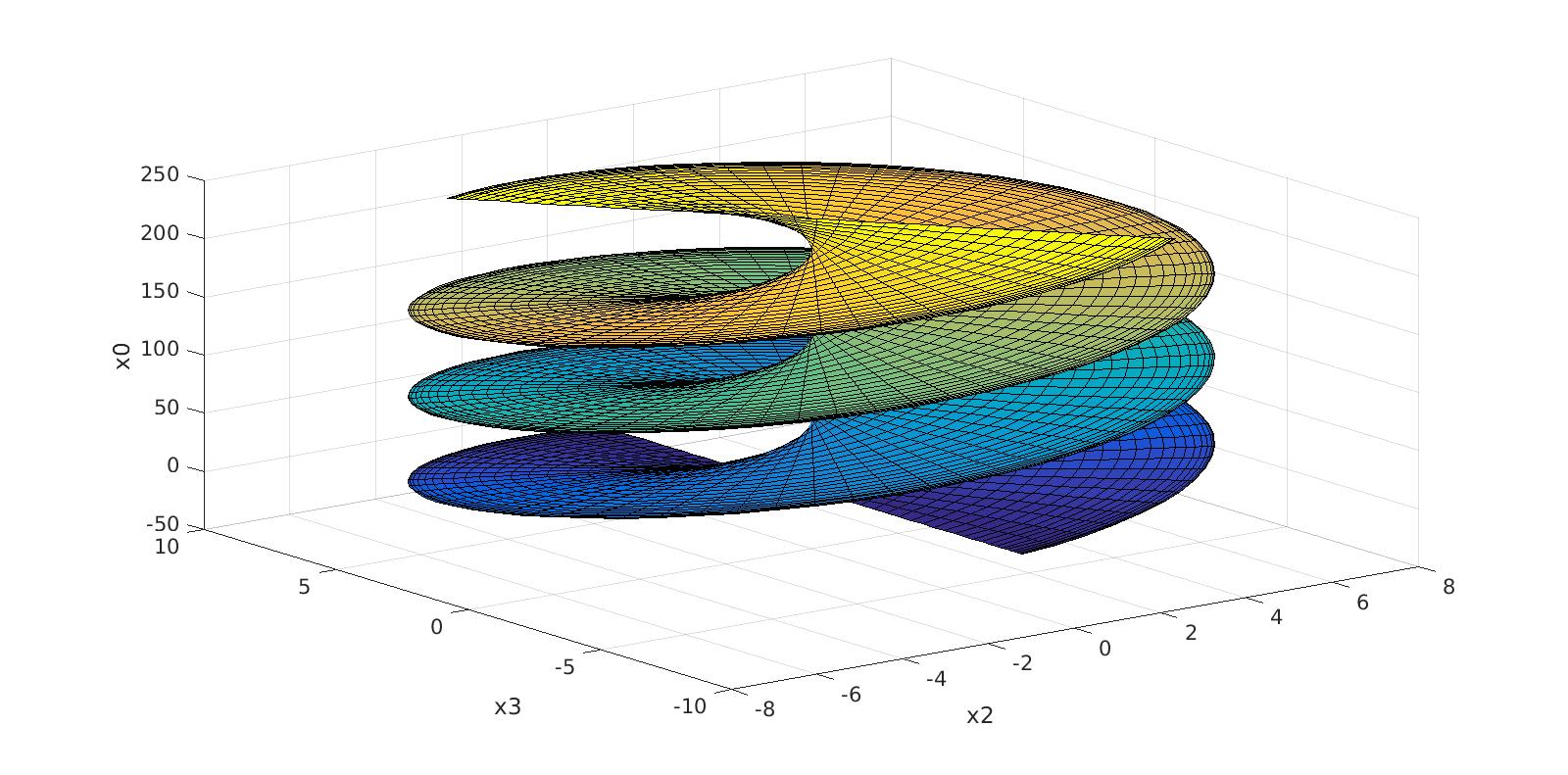}
    \caption{D = 4, rotating static string solution}
    \label{fig:4rotator}
\end{figure}

Figure \ref{fig:4rotator} was generated using a sample derivation from \cite{krijn_mode}, in order to confirm correct operation of the software for a very simple case. This is the only four-dimensional example considered in this report, and serves to demonstrate that the software is functional for any arbitrary spacetime dimensions. The solution describes a static, straight open string which rotates in the two transverse dimensions, $x^{2}$ and $x^{3}$. It is defined by the following parameters, where all transverse modes of \eqref{eq:xIdef1} not mentioned (or whose conjugates are not mentioned) are set to be 0:

\begin{equation}
\label{eq:xndef}
p^{+} = 1 ;
\qquad
x^{I}_{0} = x^{-}_{0} = 0 ;
\qquad
\alpha^{I}_1 = (4, 4i).
\end{equation}

\newpage

In the above figure and all those to follow, $x^0$ denotes time.

\begin{figure}[h!]
    \centering
    \includegraphics[width=0.5\textwidth]{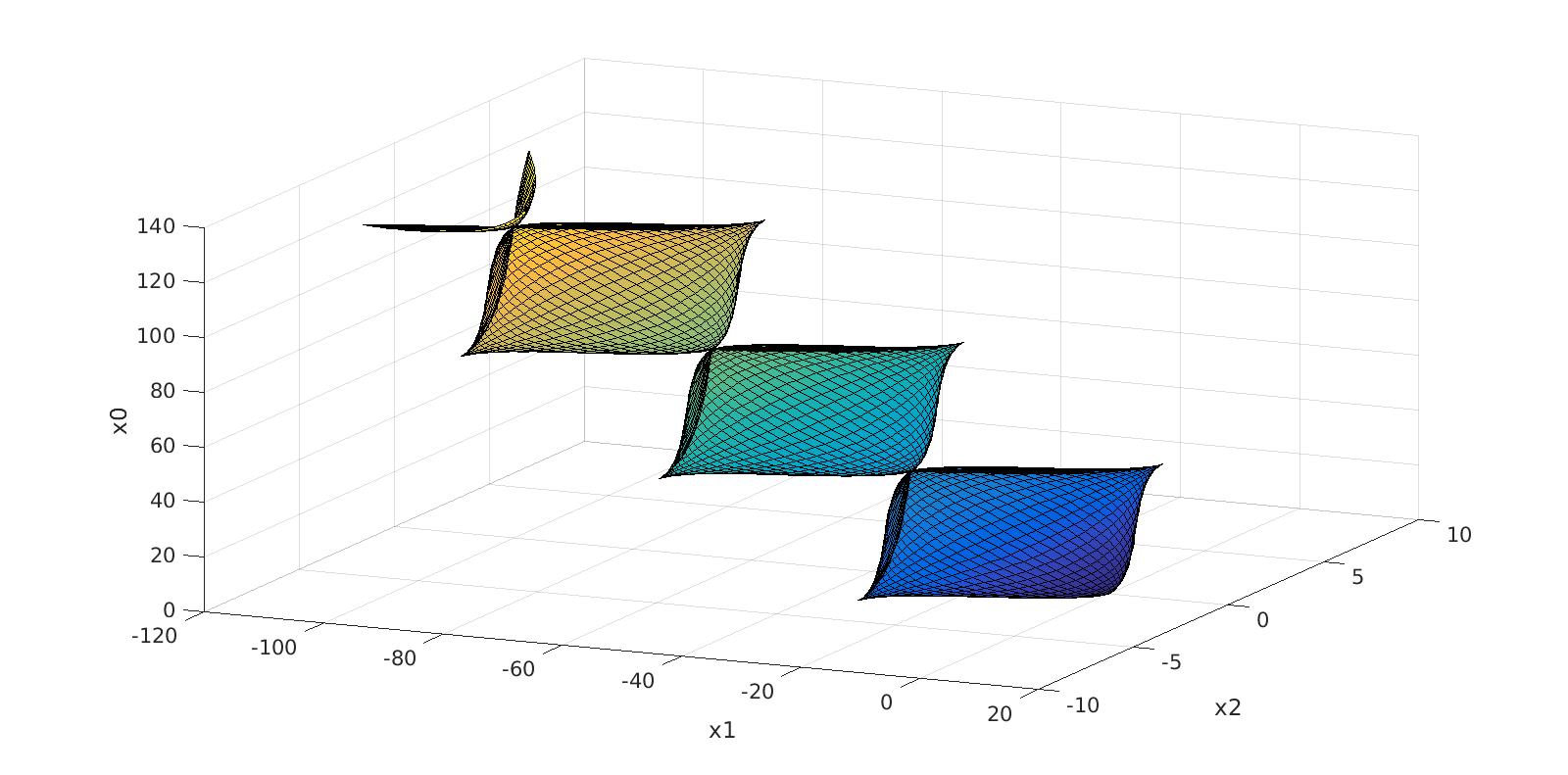}
    \caption{D = 3, single transverse-mode string solution}
    \label{fig:3straight}
\end{figure}

Figure \ref{fig:3straight} shows one of the most fundamental strings as described using the light-cone mode expansion, with only the first transverse mode being nonzero. It is defined by the following parameters:

\begin{equation}
\label{eq:xndef}
p^{+} = 1 ;
\qquad
x^{I}_{0} = x^{-}_{0} = 0 ;
\qquad
\alpha^{I}_1 = 4.
\end{equation}

\begin{figure}[h!]
    \centering
    \includegraphics[width=0.5\textwidth]{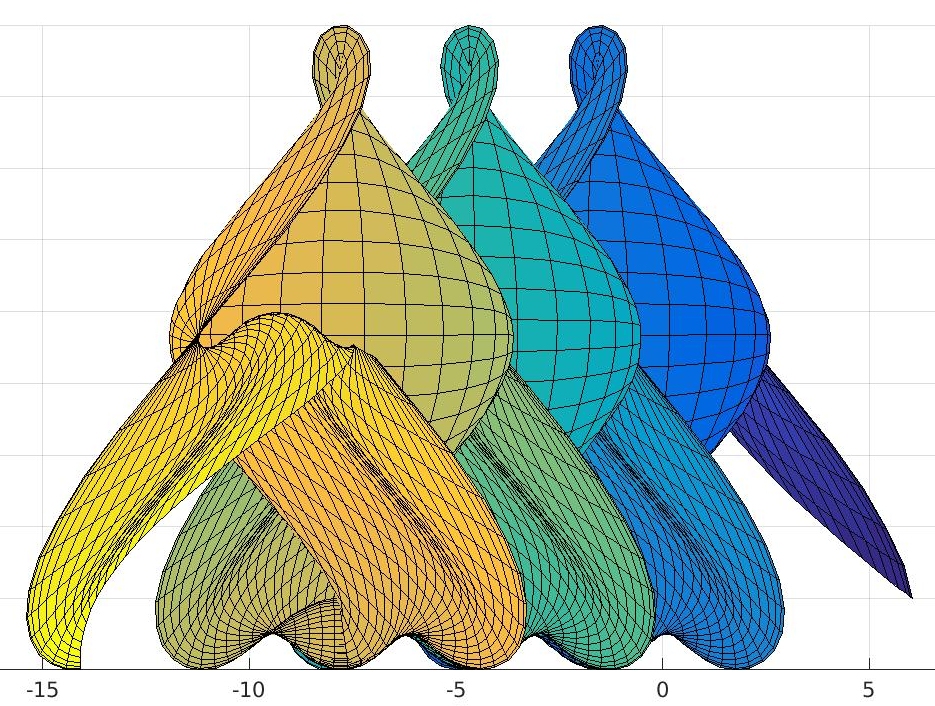}
    \caption{D = 3, asymmetric string solution, top view}
    \label{fig:asymm3}
\end{figure}

Figure \ref{fig:asymm3} shows a solution which is enriching because of its many asymmetries despite being composed of only two transverse mode oscillations. Its aesthetic beauty is also noted to be striking. This solution is defined by the following parameters:

\begin{equation}
\label{eq:xndef}
p^{+} = 1 ;
\qquad
x^{I}_{0} = x^{-}_{0} = 0 ;
\qquad
\alpha^{I}_1 = 4,
\qquad
\alpha^{I}_2 = 4i.
\end{equation}


\newpage
\subsection{Parameter Analysis}

After examining some arbitrary but simple examples, the string solution input parameters were further examined for their correlation with the shape properties of the worldsheet. The most trivial of these were the initial position parameters: $x^{-}_{0}$ and $x^{I}_{0}$, which impact only the translated position of the worldsheet. 

A much more intriguing parameter is $p^{+}$, referred to in \cite{zwiebach} as the light-cone energy. Simulations with LEGen showed a qualitative correlation between $p^{+}$ and the "speed" of the string, visibly recognizable in many of the above plots as a linear slope between the $x^{1}$ and $x^{0}$ coordinate directions. To investigate this further, the momentum slope p' was defined as:

\begin{equation}
\label{eq:slopedef}
p' = \frac{p^0}{p^1} = \frac{p^{+}+p^{-}}{p^{+}-p{-}}.
\end{equation}

In the light-cone gauge, $p^{-}$ is not treated as an independent parameter, as it is dependent on the $p^{+}$ and the zeroth Virasaro mode by the expression:

\begin{equation}
\label{eq:slopesolve}
2p^{+}p^{-} = \frac{1}{\alpha'}L^{\perp}_{0}.
\end{equation}

\begin{figure}[h!]
    \centering
    \includegraphics[width=0.5\textwidth]{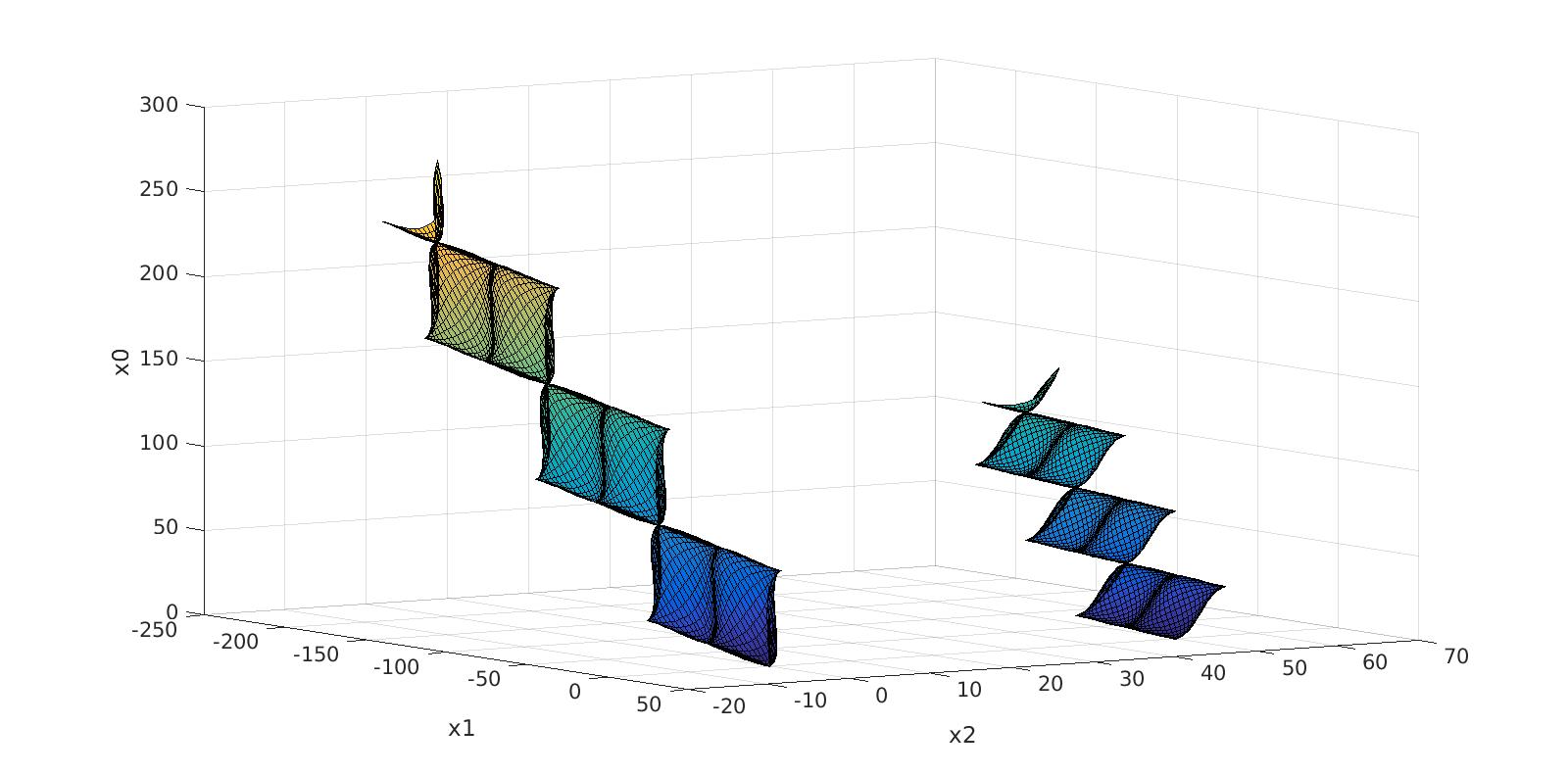}
    \caption{D = 3, double-mode solution, with $p^{+}$ varied}
    \label{fig:p_pvary}
\end{figure}

The two plots shown in Figure \ref{fig:p_pvary} are the same bimodal string solution, with only the value of $p^{+}$ varied (set to 1 on the left and 2 on the right). The right plot was translated for visual comparison effect; otherwise the worldsheets would have identical initial positions. 

It was postulated that by further increasing $p^{+}$, a string solution must be able to be generated which "points" directly in the $x^0$ direction. This may be more physically interpreted as being a string which does not translate over time, but statically oscillates. Repeated guesses of $p^{+}$, with full resimulations, seemed to place the desired value near 5, but no worldsheet produced was precisely vertical.

\begin{figure}[h!]
    \centering
    \includegraphics[width=0.5\textwidth]{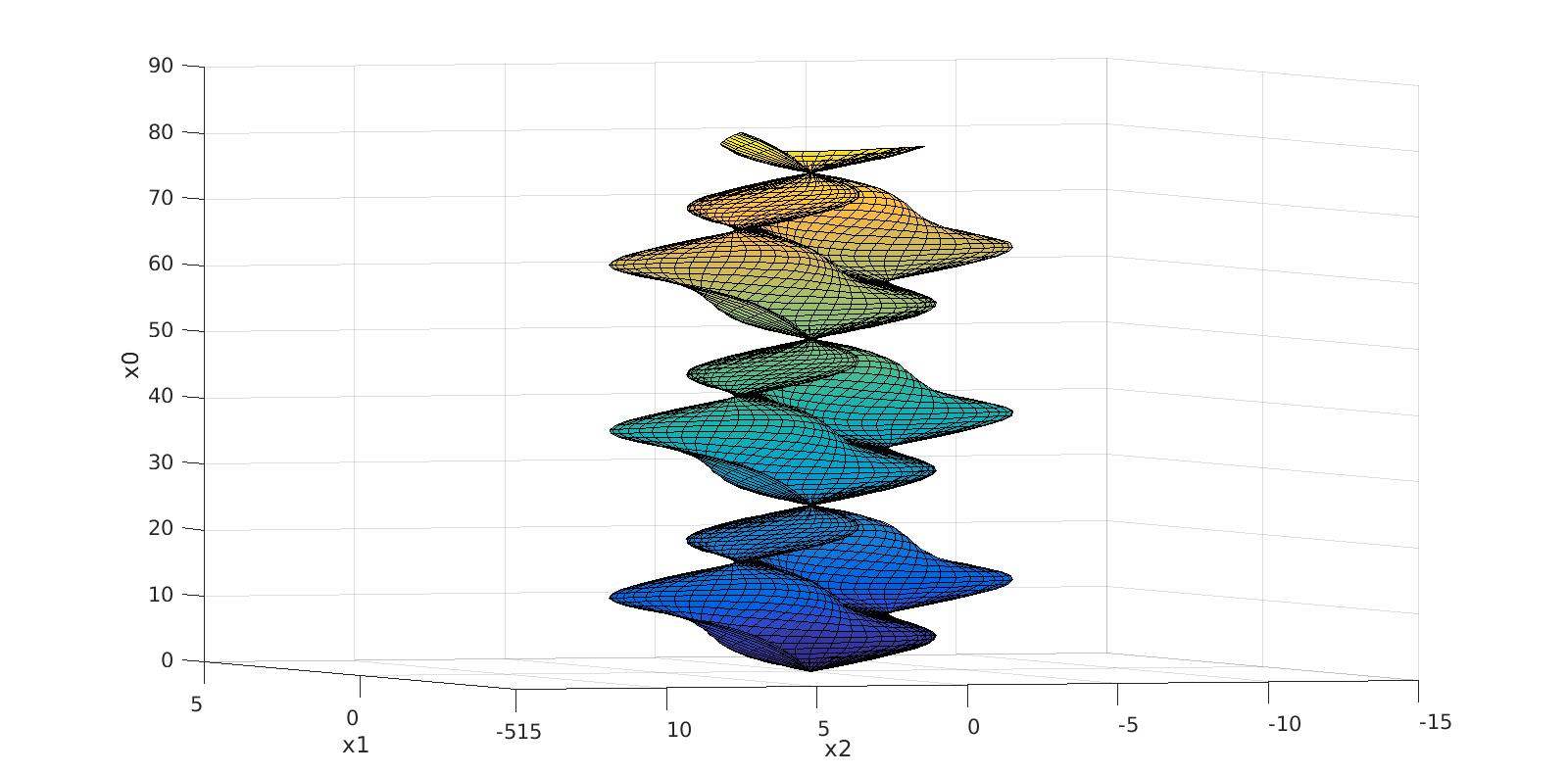}
    \caption{D = 3, non-translating single-mode solution}
    \label{fig:vert}
\end{figure}

By setting p' to a desired value of $\infty$, corresponding to vertical slope, and substituting \eqref{eq:slopedef} into \eqref{eq:slopesolve}, the necessary value of $p^{+}$ was found to be $\sqrt{32}$. A plot of the resultant worldsheet, which was indeed a static oscillating string, is given in Figure \ref{fig:vert}.

\newpage

\begin{figure}[h!]
    \centering
    \includegraphics[width=0.5\textwidth]{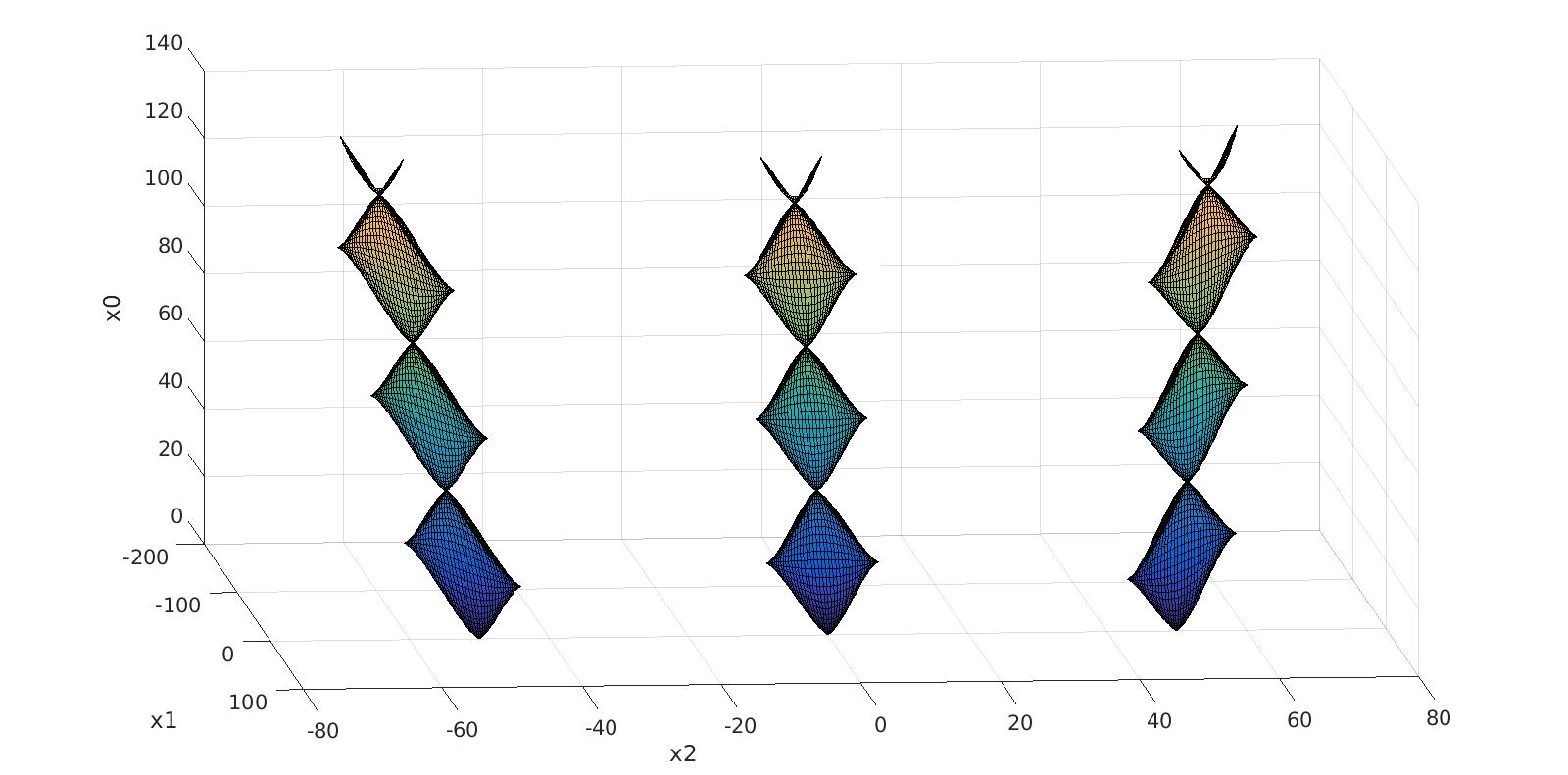}
    \caption{D = 3, single-mode solution, with $a^{I}_0$ varied}
    \label{fig:a0vary}
\end{figure}

The zero-transverse mode $\alpha^{I}_0$ was also observed to have a characteristic effect upon worldsheet shape. This parameter is constrained to be real, due to the reality condition upon \eqref{eq:xIdef1}, in which it contributes to a term linearly with $\tau$. Indeed, from this term, it was predicted that the value of $\alpha^{I}_0$ would correlate with the "slope" of the worldsheet in the transverse directions perpendicular to its overall momentum. The three plots shown in Figure \ref{fig:a0vary} are the same unimodal string solution, with only the value of $\alpha^{I}_0$ varied (set to -1 on the left, 0 in the middle, and 1 on the right). Viewed from the side (along x2), these worldsheets align perfectly, and so the observed leaning occurs only in x2 and x0.

\begin{figure}[h!]
    \centering
    \includegraphics[width=0.5\textwidth]{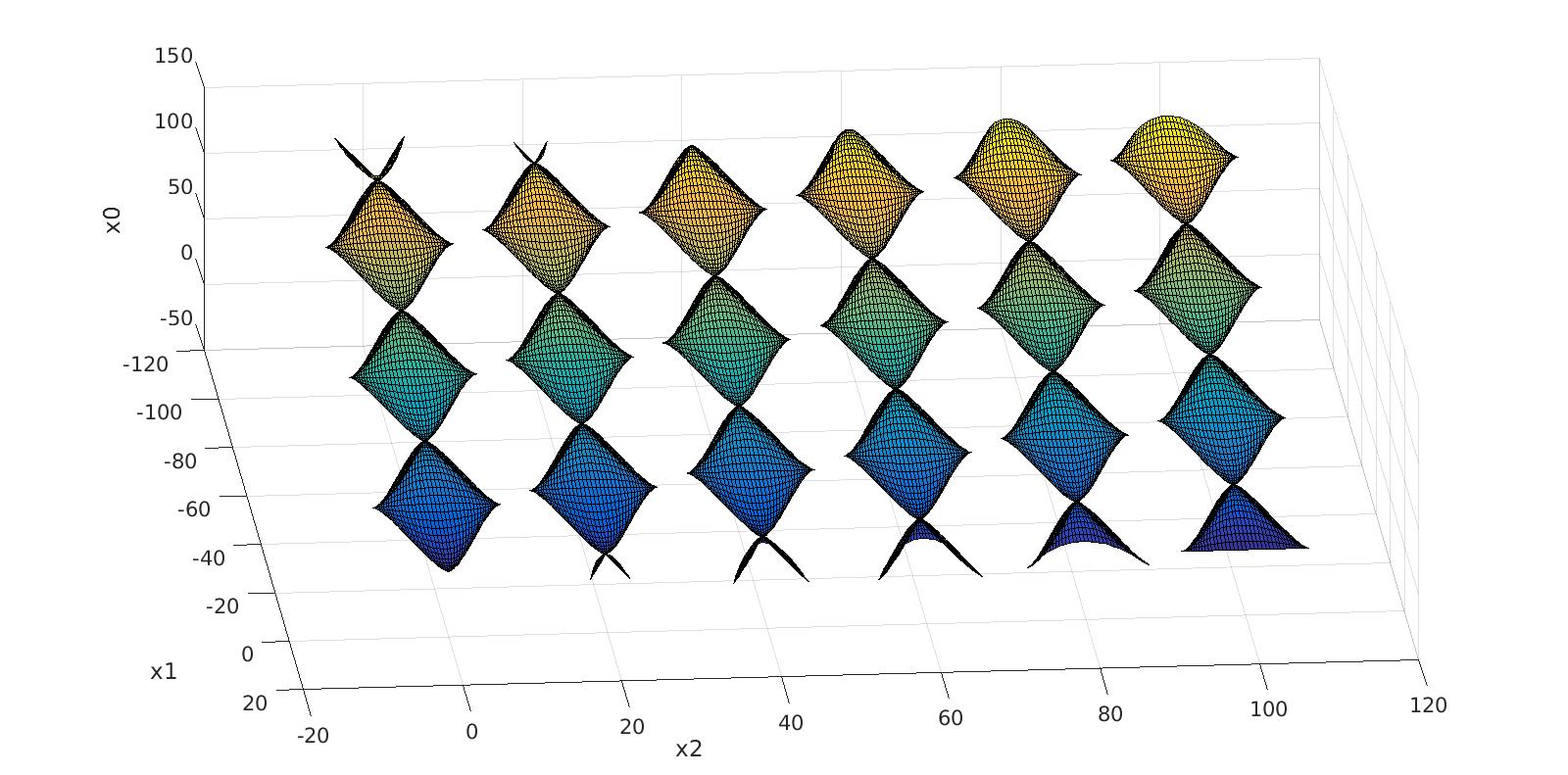}
    \caption{D = 3, single-mode solution, with $\angle a^{I}_1$ varied}
    \label{fig:phasevary}
\end{figure}

Lastly, the effect of complex rotation of a single mode coefficient was observed. The six plots shown in Figure \ref{fig:phasevary} are the same unimodal string solution, with only the value of $\alpha^{I}_1$ varied (set to 4 on the left, and rotated in the complex plane by $\frac{\pi}{10}$ for each subsequent plot). It is clear that this complex angle corresponds to the phase angle of the string solution; for the rightmost plot, a mode coefficient of $4i$ corresponds to exactly a quarter-wavelength phase shift, as would be expected.

\section{\label{sec:level1}Conclusion}

In this paper, we present the development of the  LEGen software using the light-cone gauge mode expansion technique for visualization of string solutions. Test outputs of LEGen were evaluated as proof of valid operation, and were analyzed for the properties of their input parameters, the coefficients of the mode expansion. Prospects for further development of LEGen include extending it for the visualization of superstring solutions, as well as strings winding around compact dimensions. The examples detailed in Section III.B serve only as an overview of LEGen's capabilities, which are demonstrated more fully in Part II of this work.

\subsection{Acknowledgements}

We wish to thank Professor Matthew Kleban (NYU), and Logan Boyd for helpful discussions and suggestions.

\subsection{About the Authors}

Neil Comins earned his undergraduate degree at Cornell University and his PhD at University College Cardiff, Cardiff, Wales (now Cardiff University). Graham Van Goffrier earned his undergraduate degree at the University of Maine and will enter MASt study at the University of Cambridge, Cambridge, England in Fall 2018.


\medskip
 
\bibliography{part1_publish}






\end{document}